\documentclass[12pt,epsf]{article}

\begin{document}

\begin{titlepage}

\begin{flushright}
IUHET-466\\
hep-th/0402036
\end{flushright}
\vskip 2.5cm

\begin{center}
{\Large \bf Non-Hermitian Interactions Between Harmonic Oscillators,
with Applications to Stable, Lorentz-Violating QED}
\end{center}

\vspace{1ex}

\begin{center}
{\large B. Altschul\footnote{{\tt baltschu@indiana.edu}}}

\vspace{5mm}
{\sl Department of Physics} \\
{\sl Indiana University} \\
{\sl Bloomington, IN, 47401 USA} \\

\end{center}

\vspace{2.5ex}

\medskip

\centerline {\bf Abstract}

\bigskip

We examine a new application of the Holstein-Primakoff realization of the
simple harmonic oscillator Hamiltonian. This involves the use of
infinite-dimensional representations of the Lie algebra $su(2)$.
The representations contain nonstandard raising and lowering operators, which are
nonlinearly related to
the standard $a^{\dag}$ and $a$. The new operators also give rise to a natural
family of two-oscillator couplings. These nonlinear couplings are not
generally self-adjoint,
but their low-energy limits are self-adjoint, exactly solvable, and stable.
We discuss
the structure of a theory involving these couplings. Such a theory might have
as its ultra-low-energy limit a Lorentz-violating Abelian gauge theory, and
we discuss the extremely strong astrophysical constraints on such a model.

\bigskip 

\end{titlepage}

\newpage

\section{Introduction}

The simple harmonic oscillator is one of the best understood systems in quantum
mechanics, with applications in essentially all areas of physics. However,
there may remain many interesting properties of this system that have not
been fully understood or elucidated (see~\cite{ref-thooft}, for instance).
In this paper, we present an example of this. This example provides a possible
way for nonlocal interactions to stabilize a Lorentz-violating modification to
the free photon sector of the standard model.

We shall show how the Holstein-Primakoff realization of the angular momentum
Lie algebra $su(2)$ may be used in
connection with the simple harmonic oscillator~\cite{ref-holstein}.
We first review the free
oscillator, considered
using the infinite-dimensional representations of this algebra. This leads
naturally to a study of new nonlinear couplings between multiple identical
harmonic oscillators. These couplings can be viewed in one of two ways.
Taken at face value, the interactions are not self-adjoint, and the energy
eigenvalues need not be real; however, the Hamiltonians
in question do possess low-energy limits which are self-adjoint. The second
possible
viewpoint would be to take the formula which defines the real eigenvalues
in the low-energy regime and extend that formula to cover the full range of
the quantum numbers. This has the advantage of ensuring unitary evolution, but
the additional complexity of the Hamiltonian is a corresponding disadvantage.

We shall describe how these new interactions may manifest themselves in a physical
theory, with particular emphasis on couplings between the two polarization
modes of the free electromagnetic field.
However they are interpreted, the interactions we shall discuss may be relevant
as part of a nonlocal, Lorentz-violating quantum field theory. Recently, there
has been a great deal of interest in the possibility that Lorentz symmetry may
not be exact in nature. A violation of this kind of fundamental symmetry
could arise as part of the novel physics of the Planck scale. Relics of this
violation would then persist even in the low-energy effective theory.
The general local Lorentz-violating standard model extension (SME) has been
developed~\cite{ref-kost1,ref-kost2,ref-kost12}, and the
stability~\cite{ref-kost3} and
renormalizability~\cite{ref-kost4} of this extension have been studied.
Lorentz violation is a very interesting area of theoretical physics, because
even superficially simple questions about its physics may have subtle and
even ambiguous answers.
For example, the study of the
gauge invariance properties of and finite radiative corrections to
Lorentz-violating field theories has proven to be a fruitful source
for new theoretical
insights~\cite{ref-coleman,ref-jackiw1,ref-victoria1,ref-kost5,ref-altschul1,
ref-altschul2}.

One significant difficulty with Lorentz-violating quantum field theories is that
they frequently exhibit problems with stability~\cite{ref-kost3,ref-jackiw2}. Yet
it has been suggested that some form of nonlocality might overcome this
problem~\cite{ref-kost3,ref-bros}. The harmonic oscillator interactions we
consider can
provide a concrete example for how this stabilization might work, if we adopt
the second interpretation of these interactions as described above. In the
limit of very low energies, the nonlocal interactions may couple together the
two polarization modes of a free photon in exactly the same way as would a
local, renormalizable operator from the SME~\cite{ref-kost1,ref-jackiw2}. However,
the nonlocality ensures
that the theory remains stable, even for very large photon numbers.
Weak forms on nonlocality have also been considered in other Lorentz-violating
contexts~\cite{ref-altschul3}.

The kinds of infinite-dimensional representations of $su(2)$ that we will
consider have also been introduced in the
context of the Dirac Coulomb problem~\cite{ref-martinez1,ref-martinez2} and
in generalizations of the
Dirac monopole~\cite{ref-nesterov}, where they provide useful
insights. Moreover,
other Lie algebra also possess infinite-dimensional representations that are not
representations of any corresponding Lie group. These representations might
be useful in the study of certain quantum-mechanical systems, through a
generalization of the techniques used
in~\cite{ref-martinez1,ref-martinez2,ref-nesterov}
or developed in this paper.

\section{Single-Oscillator Operators}

We shall begin by reviewing the Holstein-Primakoff realization of the harmonic
oscillator raising and lowering algebra.
Most frequently, when one studies the representations of $su(2)$ in connection
with quantum mechanics, one is interested only in the
finite-dimen\-sion\-al representations, which are countable and parameterized
by the total angular momentum $\ell$. One considers an operator $\vec{J}=
(J_{1},J_{2},J_{3})=\left(\frac{1}{2}J_{+}+\frac{1}{2}J_{-},\frac{1}{2i}J_{+}-
\frac{1}{2i}J_{-},J_{3}\right)$, with standard commutation relations. Beginning
from a highest weight state $|\ell\rangle$, with $J_{+}|\ell\rangle=0$,
one constructs each of the $2\ell+1$ states $|\ell-s\rangle$ by
acting on $|\ell\rangle$ with $J_{-}$ $s$ times and normalizing appropriately.

When $2\ell$ is a nonnegative integer, there are only these $2\ell+1$
states, because $J_{-}|-\ell\rangle=0$. However, for more general values
of the highest weight, the sequence of states does not terminate. Instead,
one constructs an infinite tower of equally spaced states. This tower of states
has a structure which is identical to that of the simple harmonic oscillator.

In fact, if we begin with a highest weight state $|\lambda\rangle$, with
$J_{3}|\lambda\rangle=\lambda|\lambda\rangle$ (where $2\lambda$ is not
a nonnegative integer), we may construct states
$|\lambda-n\rangle$ for all nonnegative integers $n$, using
$J_{-}|\lambda-n\rangle=\sqrt{\lambda(\lambda+1)-(\lambda-n)(\lambda-n-1)}
|\lambda-n-1\rangle$. Then the Hamiltonian $H_{\lambda}=-\omega J_{3}
+\omega\left(\lambda+\frac{1}{2}\right)$ has nondegenerate eigenvalues
$\left(n+\frac{1}{2}\right)\omega$, which are precisely the energy
eigenvalues of a
harmonic oscillator (when we set $\hbar=1$). Since a quantum-mechanical system
is entirely specified by its Hilbert space and the Hamiltonian acting on
that space, this is equivalent to an alternate description of the harmonic
oscillator. (Note that this description is completely distinct
from Schwinger's development of the angular momentum algebra in terms of
harmonic oscillator states~\cite{ref-schwinger}.)

We should point out that the state space is a representation
only of the Lie algebra $su(2)$, not of the Lie group $SU(2)$. That this
is the case should be clear from an examination of the spectrum of the
Hamiltonian $H_{\hat{n}}=
\omega\hat{n}\cdot\vec{J}$. This would be a Zeeman effect Hamiltonian
if the representation of ${\vec J}$ were finite-dimensional, with $\vec{J}$
transforming in the adjoint representation of $SU(2)$. The $SU(2)$
symmetry would then dictate that the eigenenergies should be independent of the
direction of $\hat{n}$. However, those energies are
clearly not independent of $\hat{n}$
for the infinite-dimensional representations we are now considering.
Specifically, if $\hat{n}=-\hat{e}_{3}$, then we
have a harmonic oscillator system, but if $\hat{n}=+\hat{e}_{3}$, then
the energy is not bounded from below. Therefore, the infinite-dimensional
operators cannot form a representation of the group $SU(2)$.

We shall now re-label our states, so that they match the usual harmonic
oscillator nomenclature. We make the replacement $|\lambda-n\rangle
\rightarrow|n\rangle$. Now, in addition to the standard harmonic
oscillator raising and lowering operators $a^{\dag}$ and $a$, we have
a new set of raising and lowering operators $J_{-}$ and $J_{+}$,
given by
\begin{eqnarray}
J_{-}|n\rangle & = & \sqrt{\lambda(\lambda+1)-(\lambda-n)(\lambda-n-1)}|n
+1\rangle \\
J_{+}|n\rangle & = & \sqrt{\lambda(\lambda+1)-(\lambda-n)(\lambda-n+1)}
|n-1\rangle,
\end{eqnarray}
or
\begin{eqnarray}
\label{eq-J-a}
J_{-} & = & a^{\dag}\sqrt{2\lambda-a^{\dag}a} \\
\label{eq-J+a}
J_{+} & = & a\sqrt{2\lambda-a^{\dag}a+1}.
\end{eqnarray}
The square root operators are to be interpreted as having eigenvalues
equal to the square roots of the eigenvalues of the operators inside,
with the same eigenvectors. For a mechanical oscillator,
we may further express the $J_{\pm}$
in terms of the position and momentum operators $x$ and $p$, using the
usual linear relations connecting $x$ and $p$ to $a^{\dag}$ and $a$; however,
this substitution
must be made with the understanding that the proper interpretation of the
$J_{\pm}$ operators requires the use of the discrete ``number of
quanta present'' basis of states.

The $J_{\pm}$ give us a new family of raising and lowering operators,
parameterized by
$\lambda$. These operators are distinct from the $a^{\dag}$ and $a$
for all finite $\lambda$. This is clear from the differing commutation
relations $[a,a^{\dag}]=1$ and
$[J_{+},J_{-}]=2J_{3}$. In general, for large $n$ (large in comparison
with $\lambda$ and unity), the matrix elements of
$J_{\pm}$ are larger that those of $a$ and $a^{\dag}$ by a factor of
${\cal O}(\sqrt{n})$.
However, the new operators do include the $a$ and $a^{\dag}$ as limiting
cases. As $\lambda\rightarrow\infty$, $\frac{J_{+}}{\sqrt{2\lambda}}
\rightarrow a$ and $\frac{J_{-}}{\sqrt{2\lambda}}\rightarrow a^{\dag}$.
(These limits are to be interpreted in terms of the matrix elements of the
operators involved, and all half-integral values of $\lambda$ must be avoided as
the limit is taken.)

We can see from (\ref{eq-J-a}) and (\ref{eq-J+a}) that $J_{+}\neq
J_{-}^{\dag}$ for finite, non-half-integral $\lambda$, because the square roots
in (\ref{eq-J-a}) and (\ref{eq-J+a}) may become imaginary. This means
that $J_{1}$ and $J_{2}$ are not self-adjoint, a difficulty which we glossed over
when
we discussed the Hamiltonian $H_{\hat{n}}$. The Casimir operator
$\vec{J}^{2}$ is self-adjoint, however. Moreover, in the basis
of eigenstates of $J_{3}$, $J_{+}=J_{-}
^{T}$.  Since the matrix
elements of $J_{-}$ and $J_{+}^{\dag}$ differ only by phase factors in the 
$J_{3}$ basis, there is
no ambiguity in defining the entire state space starting from the ground state.
Finally, we point out that if $\lambda$ is large and positive, then the
non-self-adjoint character of $\vec{J}$ does not become apparent unless $n$ is at
least comparable to $2\lambda$. These facts will prove important when we
discuss the coupling between two harmonic oscillators.

\section{Multiple Coupled Oscillators}

We shall now consider a novel application of this description of the harmonic
oscillator. We may determine the energy eigenvalues exactly for
certain systems of coupled identical
oscillators in which the couplings are nonlinear. Let us consider
the Hamiltonian
\begin{eqnarray}
H & = & H_{\lambda}+H_{\mu}+gH_{{\rm int}} \\
\label{eq-interaction}
H & = & \left[-\omega J_{A3}+\left(\lambda+\frac{1}{2}\right)\omega\right]
+\left[-\omega J_{B3}+\left(\mu+\frac{1}{2}\right)\omega\right]
+gH_{{\rm int}}.
\end{eqnarray}
$\vec{J}_{A}$ and $\vec{J}_{B}$ are two independent vectors of operators
of the type we have been considering, corresponding to the highest weights
$\lambda$ and $\mu$, respectively. $H_{{\rm int}}$ is an interaction, whose form
we shall discuss shortly.
This Hamiltonian has three adjustable parameters. $\lambda$ and $\mu$
determine the structure of the harmonic oscillator representations that we
are using. However, like $g$, they may be seen as parameters describing the
interaction, because we have shifted the total energy in such a way as to make
the spectrum of the free oscillators' Hamiltonian $H_{\lambda}+
H_{\mu}$ independent of both $\lambda$ and $\mu$. These two parameters
may be chosen freely, subject to the condition that neither $2\lambda$ nor
$2\mu$ is a nonnegative integer.

We shall choose an interaction $gH_{{\rm int}}$ that is similar in form to
$g\vec{J}_{A}\cdot\vec{J}_{B}$.
If $g\neq0$, $g\vec{J}_{A}\cdot\vec{J}_{B}$
is not self-adjoint; however, if $\lambda$ and $\mu$ are large and positive, this
problem will not be apparent in the vicinity of the ground state.
So we take $H_{{\rm int}}$ to agree with $\vec{J}_{A}\cdot\vec{J}_{B}$ when
$n_{A}+n_{B}+1<\min(2\lambda,2\mu)$, where $n_{A}$ and $n_{B}$ are the
principal quantum numbers of the two oscillators.
We may then consider the effects of this
interaction within this restricted (``low-energy'') regime.

We may solve the {\em restricted} Hamiltonian using the ordinary techniques for
the addition of angular momenta. The operator $\vec{J}=
\vec{J}_{A}+\vec{J}_{B}$ has highest weights of the form $\lambda+\mu-i$
for all nonnegative integers $i$, and each value of $i$ corresponds to
a single irreducible component of the representation.
The Clebsch-Gordon coefficients for these representations
can be calculated by the standard method of applying lowering operators and
using Gram-Schmidt orthonormalization. However, our primary interest is in
the energy levels.

The eigenvalues of $H_{\lambda}+H_{\mu}$ are just $(n_{A}+n_{B}+1)\omega$.
When we change the basis, to use the ``total angular momentum'' $\vec{J}$,
this part of the Hamiltonian becomes
\begin{equation}
H_{\lambda}+H_{\mu}=-\omega J_{3}+(\lambda+\mu+1)\omega.
\end{equation}
The eigenstates of the system are parameterized by the highest weight $\lambda+
\mu-i$ and by the ``number of quanta present'' (that is, the number of
applications of $J_{-}$ on the highest weight state required to produce
a given state), $n$. If we denote these states by $|i,n\rangle$, then
it is clear that $(H_{\lambda}+H_{\mu})|i,n\rangle=(i+n+1)\omega|i,n\rangle$.
Since $i$ and $n$ have the same range as $n_{A}$ and $n_{B}$ (all must be
nonnegative integers), this verifies that the free system has the same
spectrum in each basis.

The coupling term must be calculated in the $\vec{J}$ basis. This is easily
done, using
$\vec{J}_{A}\cdot\vec{J}_{B}=\frac{1}{2}\left(\vec{J}^{2}-\vec{J}_{A}^{2}-\vec
{J}_{B}^{2}\right)$. For a state of specified $i$, $\vec{J}_{A}\cdot\vec{J}
_{B}$ has the eigenvalue $\lambda\mu-\left(\lambda+\mu+\frac{1}{2}\right)i
+\frac{1}{2}i^{2}$. So the total energy is
\begin{equation}
\label{eq-Ein}
E_{i,n}=\omega(i+n+1)+g\left[\lambda\mu-\left(\lambda+\mu+\frac{1}{2}\right)i
+\frac{1}{2}i^{2}\right].
\end{equation}
This formula holds exactly in the entire low-energy subspace.

We must now turn our attention to the general definition of $H_{{\rm int}}$.
There are two natural ways to define this interaction. The first possibility is
that
$H_{{\rm int}}=\vec{J}_{A}\cdot\vec{J}_{B}$ exactly, and the fundamental
Hamiltonian is not self-adjoint. The second possibility involves a less drastic
modification of the structure of the theory. We simply take the exact
Hamiltonian to be defined by its eigenvalues, which have the form
(\ref{eq-Ein}). There is a complete set of states $\{|i,n\rangle\}$
corresponding to these eigenenergies, and for $i+n+1<\min(2\lambda,2\mu)$,
these states agree with the ones found above. The new Hilbert space is not
equivalent to the old, two-oscillator Hilbert space, but the restricted,
low-energy subspaces of these two Hilbert spaces are isomorphic. We shall
henceforth
adopt this second definition of $H_{{\rm int}}$ (although most of our statements
will concern only the low-energy subspace, where the two definitions are
equivalent).

A few words about the various parameters are now in order.
Our coupled system is stable only if $g\geq0$; if $g<0$, the energy is not
bounded below. We also see that the structure of the
energy levels depends only on $g$ and $\lambda+\mu$; the separate values of
$\lambda$ and $\mu$ only affect the zero-point energy.

This is a strongly nonlinearly coupled system.
We may recover the more usual result for the spectrum in the
presence of a linear coupling between the oscillators by
setting $\lambda=\mu$ and expressing
$a^{\dag}$ and $a$ in terms of the $\lambda\rightarrow\infty$ limits of
$J_{+}$ and $J_{-}$. (Note that as $\lambda\rightarrow\infty$, there are no
problems with the operators being self-adjoint.)
The interaction $g\left(a_{A}a^{\dag}_{B}+
a^{\dag}_{A}a_{B}\right)$ is the limit of $\frac{g}{\lambda}\left(
\vec{J}_{A}\cdot\vec{J}_{B}-J_{A3}J_{B3}\right)$ as $\lambda$
approaches infinity. The two terms $\frac{g}{\lambda}\vec{J}_{A}
\cdot\vec{J}_{B}$ and $\frac{g}{\lambda}J_{A3}J_{B3}$ each commute with
the noninteracting Hamiltonian;
and, although they do not commute with each other for finite values
of $\lambda$, they do commute in the infinite limit. We may see this
by evaluating the two operators for finite $\lambda$ in different
bases. In the $\vec{J}$ basis, the first term is diagonal, with
eigenvalues $g\left[\lambda-\left(2+\frac{1}{2\lambda}\right)i+\frac{1}{2\lambda}
i^{2}\right]$, just
as calculated above. The
second term is diagonal in the $J_{A},J_{B}$ basis, with eigenvalues
$\frac{g}{\lambda}(\lambda-n_{A})(\lambda-n_{B})$. As $\lambda
\rightarrow\infty$, these eigenvalues become $g(\lambda-2i)$
and $g(\lambda-n_{A}-n_{B})$, respectively. However, as we saw when we
discussed the noninteracting case, the total number operator $n_{A}+n_{B}=n+i$
is diagonal in both bases, so $g(\lambda-n_{A}-n_{B})=g(\lambda-n-i)$. Then the
total energy shift, which is now the difference between $g(\lambda-2i)$ and
$g(\lambda-n-i)$, becomes simply $g(n-i)$. That is, we have two decoupled
oscillators with frequencies $\omega\pm g$, which is just the usual
result.

More general interactions are also possible.
We may replace $g\vec{J}_{A}\cdot\vec{J}_{B}$
with an arbitrary function of $\vec{J}_{A}\cdot\vec{J}_{B}$. Any
such interaction will still commute with $H_{\lambda}+H_{\mu}$, and all the
same considerations will still apply. Other
generalizations are possible as well, through the use of other well-known
properties of $su(2)$. For example, we may generalize to the coupling of $N$
identical oscillators, with the identity
\begin{equation}
\sum_{1\leq C<D\leq N}\vec{J}_{C}\cdot\vec{J}_{D}=\frac{1}{2}\left[\left(
\sum_{C=1}^{N}\vec{J}_{C}
\right)^{2}-\sum_{C=1}^{N}\left(\vec{J}_{C}\right)^{2}\right].
\end{equation}
However, it is important to keep in mind that many
approximations that are typically used when one studies more than two
interacting angular momenta will break down when working with
infinite-dimensional representations of $su(2)$.

\section{Application to Lorentz-Violating QED}

One of the
simplest situations in which pairs of identical harmonic oscillators arise is
in an Abelian gauge theory in 3+1 dimensions. An interaction of the form
(\ref{eq-interaction}) might be relevant as a modification of the photon
sector of quantum electrodynamics. For the gauge sector alone, this
interaction could be introduced separately at each value of the photon
momentum. The selection of a specific basis of polarization states for each
momentum, and the assignment of the couplings $\lambda$, $\mu$, and $g$
generally breaks Lorentz symmetry and may also break parity invariance.
Moreover, the operators $J_{\pm}$ are nonlocal,
since they involve the total energy present in a given mode of the
electromagnetic
field, and it is not possible to express this sort of interaction conveniently
in terms of the ordinary electromagnetic field operators $A^{\mu}$ and
$F^{\mu\nu}$.
However, if we are willing to allow these modifications to the structure
of the theory, the interaction is (in the absence of charges) exactly solvable.
Since there is current interest in exotic modifications of
QED, this type
of interaction may be worthy of further investigation.

We must also say a word about the gauge invariance of this QED modification.
The $\vec{J}_{A}\cdot\vec{J}_{B}$
interaction is formulated in terms of creation and annihilation operators
(i.e.\ in the canonical quantization formalism), and so any discussion of
gauge invariance will necessarily be complicated by the difficulties that are
associated with the canonical quantization of gauge fields. However, at low
energies, the
interactions we have introduced are clearly consistent with gauge
invariance in the following sense. If we specialize to the Coulomb gauge and
quantize the transverse modes of the theory, then the low-energy interaction may
be introduced without difficulty. It does not affect the number of polarization
states (and is, in fact, strongly dependent upon this number), so it does not
spoil gauge invariance in this fashion. However, it is possible that it may
damage gauge invariance at higher energies or when interactions with matter are
considered.

Although the interactions we are considering cannot be expressed simply in terms
of the standard electromagnetic field operators, the ultra-low-energy limit of
our
theory could well be expressible in such a form. We shall shortly show that
this is indeed the case for a particular class of models. The embedding
of a ultra-low-energy
Lorentz-violating effective field theory within the framework of
our fundamental
theory is attractive for several reasons. First, the interactions we have
considered have eigenenergies that are exactly known. Second, although
Lorentz-violating theories may
exhibit stability problems, our theory does not;
the $i^{2}$ term in (\ref{eq-Ein}) ensures this. This example shows that there
can exist stable nonlocal interactions which have local, Lorentz-violating
Lagrangian theories as their ultra-low-energy limits. Third, such an embedding
demonstrates the
intriguing possibility that our conventional basis of polarization states many
be inadequate for the description of the full Hilbert space of a more
fundamental theory.

We shall therefore examine the ultra-low-energy behavior of our QED
modification, to see to what sort of effective theory it may correspond.
Since the energy-level differences depend only on $\lambda+\mu$, we shall set
$\lambda=\mu$. We take $\lambda$ to be a very large number, large in comparison
with any relevant photon occupation number. (This is what we mean by
``ultra-low-energy.'') We may then neglect $a^{\dag}a$
compared to $\lambda$ in (\ref{eq-J-a}) and (\ref{eq-J+a}). It immediately
follows that the energy eigenstates are approximately given by
\begin{equation}
|i,n\rangle\approx\frac{\left(a_{A}^{\dag}-a_{B}^{\dag}\right)^{i}
\left(a_{A}^{\dag}+a_{B}^{\dag}\right)^{n}}{\sqrt{2^{i+n}(i!)(n!)}}|0,0\rangle.
\end{equation}
Deviations from this expression are suppressed by a factor of ${\cal O}(\lambda
^{-1/2})$. The corresponding energies, in the same approximation, are
\begin{equation}
E_{i,n}\approx(n+i)\omega+2\lambda gi,
\end{equation}
where we have dropped the zero-point energy.

These results imply that the polarization modes corresponding to $a_{+}^{\dag}
\equiv\frac{1}
{\sqrt{2}}\left(a_{A}^{\dag}+a_{B}^{\dag}\right)$ and $a_{-}^{\dag}\equiv
\frac{1}{\sqrt{2}}
\left(a_{A}^{\dag}-a_{B}^{\dag}\right)$ have different frequencies. The two
frequencies are shifted from their mean value by $\pm\frac{\Delta\omega}{2}
\approx\pm g\lambda$, and the effective Hamiltonian is
\begin{equation}
\label{eq-omegaH}
H_{{\rm eff}}=(\omega+2\lambda g)a_{-}^{\dag}a_{-}+\omega a_{+}^{\dag}a_{+}.
\end{equation}
This same effective Hamiltonian arises naturally in the context of a
CPT-even, Lorentz-violating modification of the photon sector. For a theory with
Lagrange density
\begin{equation}
\label{eq-L}
{\cal L}=-\frac{1}{4}F_{\mu\nu}F^{\mu\nu}-\frac{1}{4}k_{\kappa\tau\mu\nu}
F^{\kappa\tau}F^{\mu\nu}
\end{equation}
(where $k_{\kappa\tau\mu\nu}$ has the symmetries of a Riemann tensor and
is double traceless),
the expressions for the photon modes' frequencies are (to leading order in
$k_{\kappa\tau\mu\nu}$)~\cite{ref-kost3,ref-kost4,ref-kost5}
\begin{equation}
\label{eq-omegaL}
\omega_{\pm}=(1+\rho\pm\sigma)|\vec{p}|.
\end{equation}
Here, $\vec{p}$ is the photons' 3-momentum, $\rho=-\frac{1}{2}\tilde{k}_{\alpha}
\,^{\alpha}$, and $\sigma^{2}=\frac{1}{2}\tilde{k}_{\alpha\beta}\tilde{k}
^{\alpha\beta}-\rho^{2}$, with $\tilde{k}_{\alpha\beta}=k_{\alpha\mu\beta\nu}
\hat{p}^{\mu}\hat{p}^{\nu}$ and $\hat{p}^{\mu}=(1,\vec{p}/|\vec{p}|)$. The
approximate
frequencies given by (\ref{eq-omegaL}) and (\ref{eq-omegaH}) correspond if
$\rho=\sigma=g\lambda$. So any theory with $\rho=\sigma\geq0$ for all $\vec{p}$
will
reproduce the entire low-energy behavior of our modified theory. Theories with
this property indeed exist; for example, if $k_{\kappa\tau\mu\nu}$ has the form
\begin{equation}
k_{\kappa\tau\mu\nu}=\left(v_{\kappa}u_{\tau}-v_{\tau}u_{\kappa}
\right)\left(v_{\mu}u_{\nu}-v_{\nu}u_{\mu}\right)-\frac{1}{6}\left[v^{2}u^{2}-
\left(v\cdot u\right)^{2}\right]\left(g_{\kappa\mu}g_{\tau\nu}-g_{\kappa\tau}
g_{\mu\nu}\right)+\left(\kappa\leftrightarrow\mu\right),
\end{equation}
then $\rho=\sigma=-w^{2}$, where $w^{\mu}=v^{\mu}(u\cdot\hat{p})-u^{\mu}(v\cdot
\hat{p})$. If $v^{\mu}=\left(V,\vec{0}\right)$ and $u^{\mu}=\left(0,\vec{u}
\right)$, with $|\vec{u}|
=1$, then $\rho=V^{2}\sin^{2}\theta$, where $\theta$ is the angle between
$\vec{u}$ and $\vec{p}$. This particular model is parity-preserving, with
three independent parameters.

Moreover,
{\em any} theory with nonvanishing $\sigma$ will demonstrate a splitting between
polarization modes, as would arise in our $su(2)$-modified QED. There is
therefore a large theoretical parameter space in which the theory given by
(\ref{eq-L}) can be embedded in a $su(2)$ coupling model.

These embeddings all require that, for a fixed direction $\hat{p}$, the
coupling $g\lambda$ must be proportional to $|\vec{p}|$. It would seem most
natural for $\lambda$, which represents the number of photons that must be
present in a mode of the electromagnetic field in order for the failure of the
polarization state basis to be apparent, to remain large for all values of
$|\vec{p}|$. We therefore speculate that $\lambda$ may be a
$|\vec{p}|$-independent (or $\vec{p}$-independent) constant, while $g$
scales with the magnitude of $\vec{p}$.

Based upon astrophysical experiments, the physical value of $\sigma$ is strongly
constrained, to parts in $10^{31}$ or better~\cite{ref-kost11,ref-kost14}. This
represents
an even stronger constraint on the $su(2)$ model, because $\lambda$ is
necessarily very large. It follows that $g=\sigma/\lambda$ is correspondingly
smaller. Our modification of QED is thus physically reasonable only in a very,
very small region of parameter space.

\section{Conclusions}

In summary, we have studied an applicaiotn of
the Holstein-Primakoff realization of the
simple harmonic oscillator operator algebra. This has
involved the introduction of a set of raising and lowering operators that
obey the angular momentum commutation relations.
The properties of these operators allow us to solve exactly for the low-energy
behavior of a theory with a particular nonlinear interaction. The
high-energy extension of this model
may involve either a non-self-adjoint Hamiltonian or
a new basis of states.

This model has an interesting application in Lorentz-violating physics, where
stability is typically a problem. The ultra-low-energy limit of the interaction
we have considered resembles the effect of a  small Lorentz-violating correction
to free QED. At higher energies, the nonlocality of the special harmonic
oscillator interactions we are considering can stabilize the Lorentz-violating
theory. Physically, these kinds of effects are extremely strongly constrained by
astronomical observations. However, this theory provides an useful insight into
how nonlocality and Lorentz violation may combine to form a well-behaved theory.

\section*{Acknowledgments}

The author is grateful to V. A. Kosteleck\'{y}, R. Jackiw, and M. Berger for
many helpful discussions.
This work is supported in part by funds provided by the U. S.
Department of Energy (D.O.E.) under cooperative research agreement
DE-FG02-91ER40661.

\end{document}